\documentclass[12pt]{article}
\usepackage{graphicx}
\input{epsf}
\usepackage{lineno}

\begin{document}

\title{Physics potential for the measurement of {\large ${\sigma(HZ)\times BR(H \rightarrow WW^{*})}$} at the 250 GeV ILC}
\author{ Mila Pandurovi\'c\thanks{E-mail:milap@vinca.rs}\\
{\footnotesize On behalf of the ILD concept group}\\
{\small Vinca Institute of Nuclear Science, Mihajla Petrovica Alasa 12-14. Belgrade, Serbia}\\
}
\date{}
\maketitle
\begin{abstract}

The potential of measurement of the relative statistical uncertainty of the ${σ(HZ)\times BR(H \rightarrow WW^{*})}$ decay, at the International Linear Collider (ILC) has been presented. The study is performed at the lowest energy stage of the proposed staged ILC, the 250 GeV stage. Monte Carlo samples representing all SM processes were passed through full detector
simulation  of the International Large Detector (ILD) model. Fully hadronic final state has been analyzed using an integrated luminosity of 500 fb$^{-1}$, using multivariate analysis technique. The obtained relative statistical uncertainty $\Delta (\sigma \cdot BR) \slash (\sigma\cdot BR)$ of the $\sigma(HZ)\cdot BR(H\rightarrow WW^*) $ is 4.1$\%$.\\

Talk presented at the International Workshop on Future Linear Colliders (LCWS2018), Arlington, Texas, 22-26 October 2018. C18-10-22.

\end{abstract}

\section{Introduction}

After the discovery of the Higgs boson, the measurements obtained at the LHC, in every respect, are confirmed to be in the agreement with the Standard model predictions. However, both the experimental and theoretical uncertainties leave room for the presence of new physics \cite{Peskin}, since the foreseen differences between the measured values to those of the Standard model are, in many cases, below current precision.
Therefore, the possible path for the search of the new physics is to reveal the deviations in the measurement of the observables that are sensitive to the presence of these phenomena.
One of the proposed high precision machines is the electron-positron collider ILC (International Linear Collider).
The important part of the physics program at ILC are the precise measurements of the Higgs boson properties, at the first place the measurements of its couplings.
The Standard model predicts strict linear dependence as the function of masses of corresponding particles.
The shape and the nature of the eventual deviations of the couplings can indicate the type of model for new physics 
that might be responsible.
The precision of the coupling measurements of an order of few percent is needed to be sensitive to these deviations \cite{Peskin}. \\

The ILC physics case of 250 GeV has been elaborated in detail \cite{250GeV} and  the comprehensive set of studies of the foreseen physics program have been performed in full simulation \cite{LCWS}. It was shown that the necessary sensitivity can be successfully achieved at ILC collider.\\

The talk presents the results of the measurement of the Higgs decay into a pair of $W$ bosons, at the nominal center-of-mass energy, 250 GeV of ILC, using Higgstrahlung as Higgs production process. The relative statistical accuracy of the measurement of the $\sigma (HZ) \times BR (H \rightarrow WW ^ {*}) $, allowing the determination of the precision of $\frac{{g_{HZZ}^2}{g_{HWW}}^2}{\Gamma}$.

\section{Simulation and analysis tools}

A common linear collider software package, iLCSoft \cite{ILCSoft}, is used. Signal and background samples are simulated using the Whizard 1.95 \cite{Whizard} event generator, including initial state radiation and a realistic luminosity spectrum. The hadronisation and fragmentation of the Higgs and vector bosons were simulated using Pythia 6.4 \cite{Pythia}. For the completeness, $\gamma\gamma$ to hadrons background, although negligible at this energy stage, is overlaid over each generated event sample before reconstruction.
Particle reconstruction and identification was done using the particle flow technique, implemented in the Pandora particle-flow algorithm (PFA) \cite{PFA1,PFA2}. The ILD$\_$o1$\_$v05 detector model was used.

\section{Event samples}

\begin{figure}[t]
\centering
\includegraphics[width=6.5cm]{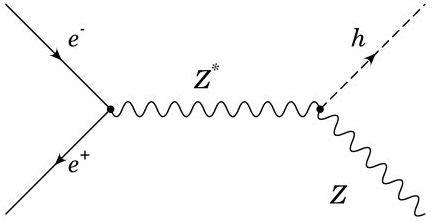}
\caption {The Feynman diagram of the dominant Higgs production processes (Higgsstrahlung).}
\label{fig:HZ}
\end{figure}

The staged concept of ILC \cite{ILCStaged}, foresees the first stage at $\sqrt{s}$ = 250 GeV. The dominant Higgs production channel is the Higgsstrahlung, with the cross section of 346 fb, including the maximal left beam polarization of (e$^{-}$,e$^{+}$) = (-80$\%$, +30$\%$). The Feynman diagram of the Higgsstrahlung Higgs production channel is shown in the Figure \ref{fig:HZ}.\\
The analyzed signal events are fully hadronic, where all the bosons in the event, the $Z$ boson, as well as both $W$ bosons decay hadronically. The corresponding signal cross-section is 36.5 fb. The signal and background processes that are considered are listed in the Table \ref{Table:Background}.

\begin{table}[t]
\caption {List of considered background processes, with the corresponding cross-sections, preselection and final selection efficiency. The last column gives the number of events in the final sample. The $\sqrt{s}$ = 250 GeV with integrated luminosity of 500fb$^{-1}$ is assumed. } \label{Table:Background}
\centering
\begin{tabular}{lrrrr}

\\\hline

Process & $\sigma[fb]$ & $\epsilon_{pres}$ [$\%$]& $\epsilon_{tot}$ [$\%$]& evts$_{fin}$\\
\hline
signal & 36.5 & 89.2 & 30.0 & 5600\\
\hline
other Higgs decays & 309.8 & 54.7 & 4.1 & 6338\\
$e^{+}e^{-}\rightarrow $ 2f hadronic &129148.6 & 1.5 & $<10^{-2}$ & 5410\\
$e^{+}e^{-}\rightarrow $ 4f WW hadronic & 14874.3 & 33.5 & 0.2 &14961\\
$e^{+}e^{-}\rightarrow $ 4f WW/ZZ hadronic & 12383.3 & 33.8 & 0.2 &13340\\
$e^{+}e^{-}\rightarrow $ 4f ZZ hadronic & 1402.0 & 42.9 & 1.0 & 7178\\
$e^{+}e^{-}\rightarrow $ 4f WW semileptonic & 18781.0 & 0.5 & $< 10^{-5}$& - \\
$e^{+}e^{-}\rightarrow $ 4f ZZ semileptonic & 1422.1 & $5\cdot 10^{-4}$ & $< 10^{-2}$& 49\\

\hline
\label{tab:Table1}
\end{tabular}
\end{table}

\section{Event selection and results}

The events are clustered into six jets the k$_{T}$ clustering algorithm \cite{Kt}, the exclusive mode, using FastJet \cite{FastJet}. In addition, the events are independently forced into two jets, in order to reduce the $H\rightarrow b\bar{b}$ background. Therefore, b and c-tagging probabilities, as determined by LCFIPlus package \cite{LCFI}, are assigned to each jet in the event, for both, six-jet and additional two-jet hypothesis. In the next step, the signal process is reconstructed by the jet pairing to form candidates for the $Z$ boson and one on-shell and one off-shell $W$ boson, the latter two coming from the Higgs decay. The combination of the pairs of jets, for which the following $\chi^{2}$ has a minimum, is chosen:

$$\chi^{2}= \frac{m_{ij}-m_{W}} {{\sigma_{W}}^{2}}+ \frac{m_{kl}-m_{Z}} {{\sigma_{Z}}^{2}}+\frac{m_{ijmn}-m_{H}} {{\sigma_{H}}^{2}} $$

\noindent where m$_{ij}$, m$_{kl}$ are the invariant masses of a di-jet pairs, which are assigned to the real W and the $Z$ boson candidates, while m$_{ijmn}$ is the invariant mass of the jets that comprize the Higgs boson candidate. m$_{V}$ are the masses and $\sigma_{V}$, (V = $W$, $Z$, $H$), are the uncertainties of the corresponding boson candidates.

The jet-opening of the clustered jets was optimized according to the criteria of minimal widths and the closest reconstructed mean of the reconstructed dijet mass distributions to that of the $Z$ and the real $W$ boson. The best results are obtained with the jet opening of R=1.5.

The cross-sections of the considered background processes are several orders of magnitude higher than for the signal events (Table 1), therefore at the next step, the background to signal ratio is minimized by the set of preselection criteria prior to the final selection.

\begin{figure}
\centering
\includegraphics[width=10.5cm]{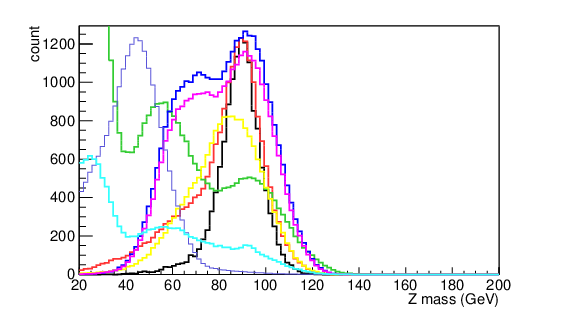}
\caption { The invariant mass of the $Z$ boson for signal (black) and background (colour). The scaling is arbitrary.}
\label{fig:MZ}
\end{figure}

\begin{figure}[h]
\centering
\includegraphics[width=10.5cm]{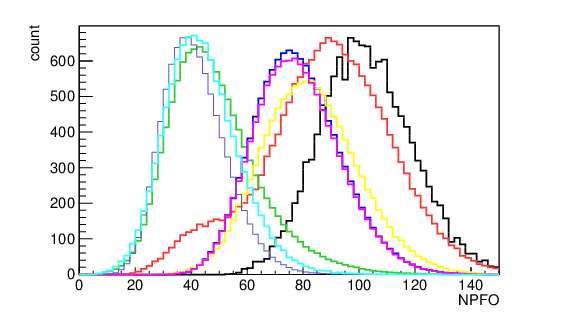}
\caption {The number of particle flow object for signal (black) and background processes (colour). The scaling is arbitrary.}
\label{fig:NPFO}
\end{figure}

The list of the preslection variables with the corresponding cut-off values:
\begin{itemize}
\item the invariant mass of the $Z$ boson candidate , $m_{Z} > 70$ GeV, ( see Figure \ref{fig:MZ}.)
\item the invariant mass of the Higgs boson candidate , $m_{H} >100$ GeV,
\item the invariant mass of the real $W$ boson candidate, $m_{W} > 60$ GeV,
\item number of particle flow objects, NPFO $>$ 70, ( see Figure \ref{fig:NPFO}.)
\item visible energy, $E_{vis} > $200 GeV,
\item transverse momentum of a single jet, $p_{T}<$20 GeV,
\item event thrust$<$0.9,
\item -log(y$_{45}) <$ 2.2, -log(y$_{56}) <$3.0, -log(y$_{45}) <$ 4.4, -log(y$_{56}) < $4.8,
\end{itemize}

\noindent where -log(y$_{ij})$ is k$_{T}$ value at which the jet number is making the transition from the i-th to the j-th number of jets in the event.

The efficiencies of reduction by the preselection are given in Table \ref{tab:Table1} for signal and background processes.

The final event selection is achieved with a multivariate analysis, using the Boosted Decision Tree (BDT) method, as inplemented in TMVA \cite{TMVA}. The set of training background samples, as well as the set of input variables, are optimized. The final event selection is performed using purely hadronic background with the following discriminating input variables:
\begin{itemize}
\item the invariant masses of both $W$ bosons, $Z$ and Higgs boson, $m_{W}$, $m_{Z}$, $m_{H}$,
\item number of particle-flow objects (NPFO) in the event,
\item transverse momentum of jets that comprize the Higgs boson, $p_{T}^{Higgs}$,
\item the highest transverse momentum of a jet in the event, ${p_{T}}^{max}$,
\item jets transitions -log($y_{12}$), -log($y_{23}$), -log($y_{34}$), -log($y_{45}$), -log($y_{56}$), -log($y_{67}$),
\item event shape variables (thrust, oblatness, sphericity and aplanarity),
\item second highest flavor tagging probabilities for the two jet hypothesis (btag, ctag),
\item angle that comprise the $Z$ boson,
\item angle that comprise the real $W$ boson.
\end{itemize}
The classifier cut value of the final selection, is chosen to minimize the relative statistical uncertainty of the cross-section $\sigma(HZ)\times BR(H\rightarrow WW^{*})$:

\begin{equation}
\frac{\Delta\sigma}{\sigma} =\frac{N_{S}}{\sqrt{(N_{S}+N_{B})}},
\end{equation}

\noindent where $N_{S}$, $N_{B}$ are the number of signal and background events after the final selection, respectively.

The number of signal and background events after the final selection are given in the Table \ref{tab:Table1}.

The obtained result of the relative statistical uncertainty of the $\sigma(HZ)\times BR(H\rightarrow WW^{*})$ measurement at 250 GeV stage of ILC is 4.1 $\%$.

\section{Conclusion}

In this talk, the measurement accuracy, $\Delta (\sigma \cdot BR) \slash (\sigma\cdot BR)$, of the Higgs decay to a $W$ pair is presented. Fully hadronic final states is considered.
The study is performed at the first energy stage of the ILC, $\sqrt{s}$ = 250 GeV, using the Higgsstrahlung Higgs production channel. The assumed beam polarization is P(e$^{-}$, e$^{+}$) = (-80$\%$, +30$\%$), with the integrated luminosity of 500 fb$^{-1}$ and the mass of Higgs boson of 125 GeV. The signal and full SM background, including beam-induced background was passed through full detector simulation and reconstruction.
\\ The obtained result for the relative statistical uncertainty of the $\sigma(HZ)\cdot BR(H\rightarrow WW^*) $ is 4.1$\%$.
This particular measurement is important for model independent determination of Higgs total width, and therefore for global fit, and it has been shown that this result improves the total Higgs decay width by of approximately 10$\%$.

\section*{Acknowledgements}
The author acknowledges the support received from the Ministry of Education, Science and Technological Development (Republic of Serbia) within the projects OI171012, and also would like to thank the LCC generator working group and the ILD software working group for providing the simulation and reconstruction tools and producing the Monte Carlo samples used in this study. This work has benefited from computing services provided by the ILC Virtual Organization, supported by the national resource  providers of the EGI Federation and the Open Science GRID.

\end{document}